# Noise Cancellation in Cognitive Radio Systems: A Performance Comparison of Evolutionary Algorithms

Adnan Quadri, Mohsen Riahi Manesh, and Naima Kaabouch
Department of Electrical Engineering,
University of North Dakota, Grand Forks, 58203, USA
E-Mail: adnan.quadri@und.edu; mohsen.riahimanesh@und.edu; naima.kaabouch@engr.und.edu

***Abstract*— Noise cancellation is one of the important signal processing functions of any communication system, as noise affects data integrity. In existing systems, traditional filters are used to cancel the noise from the received signals. These filters use fixed hardware which is capable of filtering specific frequency or a range of frequencies. However, next generation communication technologies, such as cognitive radio, will require the use of adaptive filters that can dynamically reconfigure their filtering parameters for any frequency. To this end, a few noise cancellation techniques have been proposed, including least mean squares (LMS) and its variants. However, these algorithms are susceptible to non-linear noise and fail to locate the global optimum solution for de-noising. In this paper, we investigate the efficiency of two global search optimization based algorithms, genetic algorithm and particle swarm optimization in performing noise cancellation in cognitive radio systems. These algorithms are implemented and their performances are compared to that of LMS using bit error rate and mean square error as performance evaluation metrics. Simulations are performed with additive white Gaussian noise and random nonlinear noise. Results indicate that GA and PSO perform better than LMS for the case of AWGN corrupted signal but for non-linear random noise PSO outperforms the other two algorithms.**

## I. INTRODUCTION

In communication systems, data integrity can be impacted by several factors, including noise, multipath, and shadowing. In general, sources of noise include thermal noise, noise rooting from system non-linearity in the radio front end, and interference between co-located wireless nodes within a network [1-4]. To get rid of the noise from the received signals, filters are employed in communication systems. These filters are built using hardware components, which leads to costly and bulky systems that can only filter specific frequencies [5]. However, next-generation communication technologies will host reconfigurable hardware and will enable advanced digital signal processing. Therefore, filters for these advanced systems should be programmable and should have the ability to de-noise signals of any frequency.

A promising advanced communication technology is Cognitive Radio (CR) [6-8]. A CR system operates with full-duplex communication and consists of a wideband transceiver that can configure its communication parameters according to the environment. However, these systems are impacted by additional system-induced noise sources. As a wideband transceiver, CR systems can sense multiple bands at the same time, resulting in interference-generated noise [9-10]. Similarly, in full-duplex communication, the CR receiver is saturated by noise when the co-located CR transmitter is transmitting on the same or close channel. In addition, noise from system non-linearity and thermal noise are also present in CR systems [11].

Traditional filters cannot adapt to changing frequencies and multiple bands. To that end, adaptive filters must be employed to de-noise a signal of any frequency by readjusting filter parameters during the operation. Several adaptive techniques for noise cancellation have been proposed, including search optimization algorithms [12-15]. Briefly, in an adaptive filter, the received noisy signal is subjected to filtering and the filtered output is compared against a desired signal to compute the error. The task of the adaptive algorithm is to search for an optimal solution that minimizes the error (i.e., the global minima of the error surface). Previous studies employed gradient-descent based search optimization algorithms, which initialize with a predefined guiding factor and follow the slope of the gradient to locate the desired minima of the error surface. Examples of these algorithms include least mean square (LMS) and its variants – normalized LMS (NLMS) [12], recursive least square (RLS) [13], and filtered x-LMS (FxLMS) [14]. However, these algorithms are only able to identify the local minima of a multimodal error surface and are highly dependent on the appropriate selection of their initialization variables [15]. For instance, LMS algorithm initializes with a step size variable that acts as the controlling parameter for the convergence of the algorithm. A larger step size value renders high steady state misadjustment, but smaller values decrease the convergence speed of the algorithm [16]. In addition, these gradient-descent based algorithms experience degrading performance for signals with random and non-linear noise [17].

A better alternative to gradient-descent algorithms is non-gradient algorithms, specifically evolutionary algorithms which are able to find the global minima of the error surface and can adapt to drastic changes in signals. Examples of these algorithms include genetic algorithm (GA) [17-18] and particle swarm optimization (PSO) [16]. These techniques, also referred to as global search optimization techniques, are based on evolutionary computation that mimics animal behavior and human evolution. Other than being able to locate global minima, non-gradient algorithms do not rely on a single variable initialization and are capable of adapting to random noise [19-21].

In previous work, we have implemented PSO for de-noising signals in CR systems [22]. In this paper, we investigate the efficiency of GA in dynamically filtering signals in CR systems and compare its performance to those of PSO and LMS algorithm. The paper is organized as follows. Section II includes the description of system models of GA, PSO, and LMS algorithm. In section III, results from the simulation are discussed and a general performance comparison of all the three algorithms is provided. At the end, a conclusion is drawn at section IV.

## II. Methodology

Fig. 1 shows the block diagram of the proposed system. This system was implemented using MATLAB. Stream of information bits are generated and modulated using M-ary phase shift keying (M-PSK) modulation scheme to be transmitted as signal, $x(t)$. Two cases of received noisy signal, $r(t)$, are developed in the simulation by adding noise to the transmitted signal, $x(t)$. For the first case, additive white Gaussian noise (AWGN) is added to the transmitted signal. For the second case, in addition to the AWGN, nonlinear noise is added to the signal. Received noisy signal corrupted with both AWGN and nonlinear noise is referred to as random noisy signal throughout the rest of the paper. After the addition of noise, signal $r(t)$ is sampled at the receiver radio's front end and forwarded to the adaptive filter, where it goes through the process of noise cancellation by one of the three filtering techniques.

The adaptive noise cancellation block employs an adaptive line enhancer (ALE) as the adaptive filter instead of active noise control (ANC) based filtering system. As illustrated in Fig. 2, an ALE based filtering system uses only one sensor and produces a delayed version of the received signal for noise cancellation. In ANC, a secondary reference sensor is required to estimate the noise in a noisy signal [12]. The received samples of noisy signal, $\boldsymbol{d}[n]$, is fed to the ALE, which creates a delayed version, $\widehat{\boldsymbol{y}}[n]$ of the received samples, $\boldsymbol{d}[n]$ by introducing a delay of , $Z^{-\Delta}$ . The filtered output signal, $\boldsymbol{y}[n]$ is estimated by updating weight coefficients $\boldsymbol{W}[n]$, which is supplied by the GA/PSO/LMS of the adaptive filter. The output can be expressed as:

$$\boldsymbol{y}[n] = \widehat{\boldsymbol{Y}}[n]\boldsymbol{W}[n] \quad (1)$$

$$\widehat{\boldsymbol{Y}}[n] = [\widehat{\boldsymbol{y}}[n], \widehat{\boldsymbol{y}}[n-1], \ldots, \widehat{\boldsymbol{y}}[n-L+1]] \quad (2)$$

$$\boldsymbol{W}[n] = [W_1, W_2, \ldots, W_L]^T, \quad (3)$$

where, L is the adaptive filter order and T represents the transpose of the weight vector. To find the optimal weight solution for noise cancellation, the error, difference between the received samples and filtered output, is calculated and minimized. This error signal $\boldsymbol{e}[n]$ is expressed as:

$$\boldsymbol{e}[n] = \boldsymbol{d}[n] - \boldsymbol{y}[n] \quad (4)$$

The filtered output is then processed by the analog-to-digital converter and converted to baseband received bits for the purpose of demodulation. Although the filtering algorithms, PSO and LMS, have been described in our previous work [19], an overview of the two algorithms besides GA is also provided in this paper.

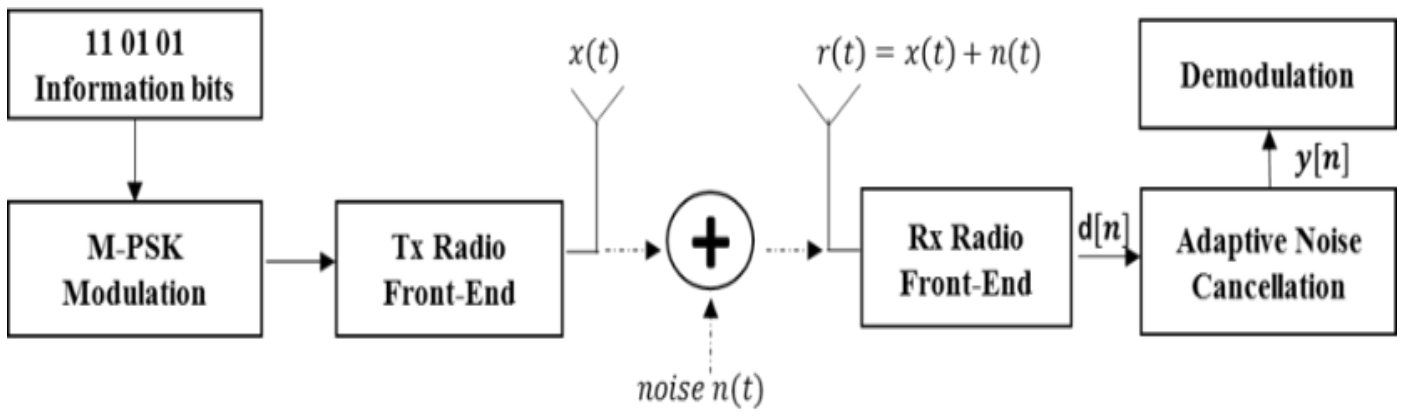

Fig. 1. System block diagram

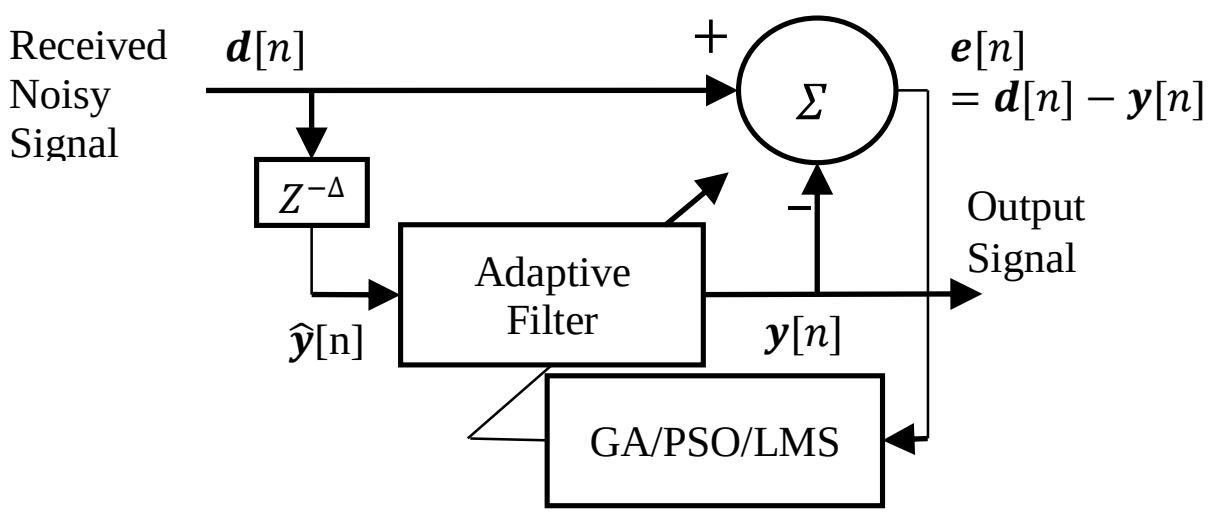

Fig. 2. ALE based adaptive filter

### A. De-noising signals using GA

Genetic Algorithm is a global search optimization technique that can locate the global minima of a multimodal error surface. This algorithm mimics the biological evolution and follows the 3-step cycle: evaluation, selection, and reproduction [17-18]. It starts with a set of population, $P_{ga}$ (also referred to as chromosomes) and this set is then evaluated for its fitness to minimize the error over $J$ generation. Once the evaluation steps are completed, the most fit chromosomes or parents are selected to mate. In the last step, the selected chromosomes bear offspring and these children are used as the next set of population or parents for the next generation until maximum number of generation is reached or the global minima is located.

Precisely, for $P_{ga}$ the number of population random solutions are generated as $P_i = [P_1, P_2, \ldots, P_L]$, where $i = 1, \ldots\ldots, P_{ga}$. The first set of solutions are then binary decoded and forwarded for fitness evaluation. The fitness or cost function is defined to minimize the error for the $i^{th}$ solution in $J^{th}$ generation. This fitness is expressed as:

$$f_{i,J} = \frac{1}{H}\sum_{n=1}^{H} \boldsymbol{e}_{i,J}[n]^2, \quad (5)$$

where $\boldsymbol{e}_{i,J}[n]$ is the error signal for the $i^{th}$ population of $J^{th}$ generation and $H$ is the input samples to the filter. After the fitness is calculated, minimum fitness is stored as best fitness and portion $\beta$ of the population $D = \beta.\ f$ parents are selected and passed into the next generation. Using the roulette wheel selection procedure $D$ parents are mated to generate children, which then undergo crossover and mutation. The mutation rate impacts the convergence of GA – a too low mutation rate within a reasonable number of generation is not sufficient for the convergence of GA, whereas a high mutation rate may cause GA to diverge [17]. Similarly, crossover introduces genetic diversity and usually is set based on engineering experiences [18]. In this paper, the crossover and mutation rates are defined as $P_c$ and $P_m$ and are set to a value for which GA renders the lowest mean square error. In this work, simulations were performed to define the mutation and crossover rates, which are shown in the results section of this paper. Once the new set of population is generated it undergoes fitness re-evaluation and the best fit portion of the population is kept. As shown in the flowchart of Fig. 3, the above mentioned processes continue until the maximum number of generation is reached or the optimal solution is found.

### *B. De-noising signals using PSO*

PSO algorithm is based on stochastic global optimization techniques. Motivated by the social interaction of bird flocking and fish swarms, PSO was proposed by James Kennedy and R.C. Eberhart in 1995 [14]. When in search for food, birds share their respective positions and update the flock with the information on the best food source within the search space. In the case of adaptive noise cancellation, similar search pattern is used in PSO with the objective of minimizing residual noise by locating best weight coefficients for the adaptive filter, which is analogous to finding the best food source or position. In order to cancel the noise, a cost function is defined that calculates the mean square error (MSE) between the received samples $\boldsymbol{d}[n]$, and the filtered output $\boldsymbol{y}[n]$.This cost function is defined as:

$$C_{i,k} = \frac{1}{H}\sum_{n=1}^{H} \boldsymbol{e}_{i,k}[n]^2, \qquad (6)$$

where $\boldsymbol{e}_{i,k}[n]$ is the error signal for the $i^{th}$ particle and $k^{th}$ iteration and $H$ is the input samples to the filter. Once the error is minimized by identifying the optimal solution or weight coefficients, PSO supplies the solution to the filter, which in turn produces the filtered output $\boldsymbol{y}[n]$, as in (1). Precisely, PSO starts by defining a set of particles and their respective velocities where the initial velocities are set to be zero. Here, the weight coefficients are represented by the position vector that is initialized as $N$ number of random solutions $\boldsymbol{w}_i = [w_1, w_2, \ldots, w_L]$, where $i = 1, \ldots N$. The cost with the first set of particle positions is then calculated for maximum of $k$ iterations and $N$ particles. When the minimum value of the cost function is attained by PSO, the respective particle position for the minimum cost is set as the best cost, $P_{bestcost}$. Over $k$ iterations, the velocity of each of the $N$ particles is updated from the initial value of zero and is defined as:

$$\boldsymbol{v}_{i,k} = \boldsymbol{v}_{i,k-1} + c_1 r_1 (\boldsymbol{P}_{bestcost,k} - \boldsymbol{w}_{i,k-1}) + c_2 r_2 (\boldsymbol{P}_{globalbest,k} - \boldsymbol{w}_{i,k-1}) \qquad (7)$$

where, $c_1, c_2$ are global and local learning coefficients, $\boldsymbol{w}_{i,k-1}, \boldsymbol{v}_{i,k}$ are the position and velocity, respectively, and $r_1, r_2$ are random numbers within the range of 0 and 1. For the $i^{th}$particle at $k^{th}$ iteration, the position of the particle is updated using:

$$\boldsymbol{w}_{i,k} = \boldsymbol{w}_{i,k-1} + \boldsymbol{v}_{i,k} \qquad (8)$$

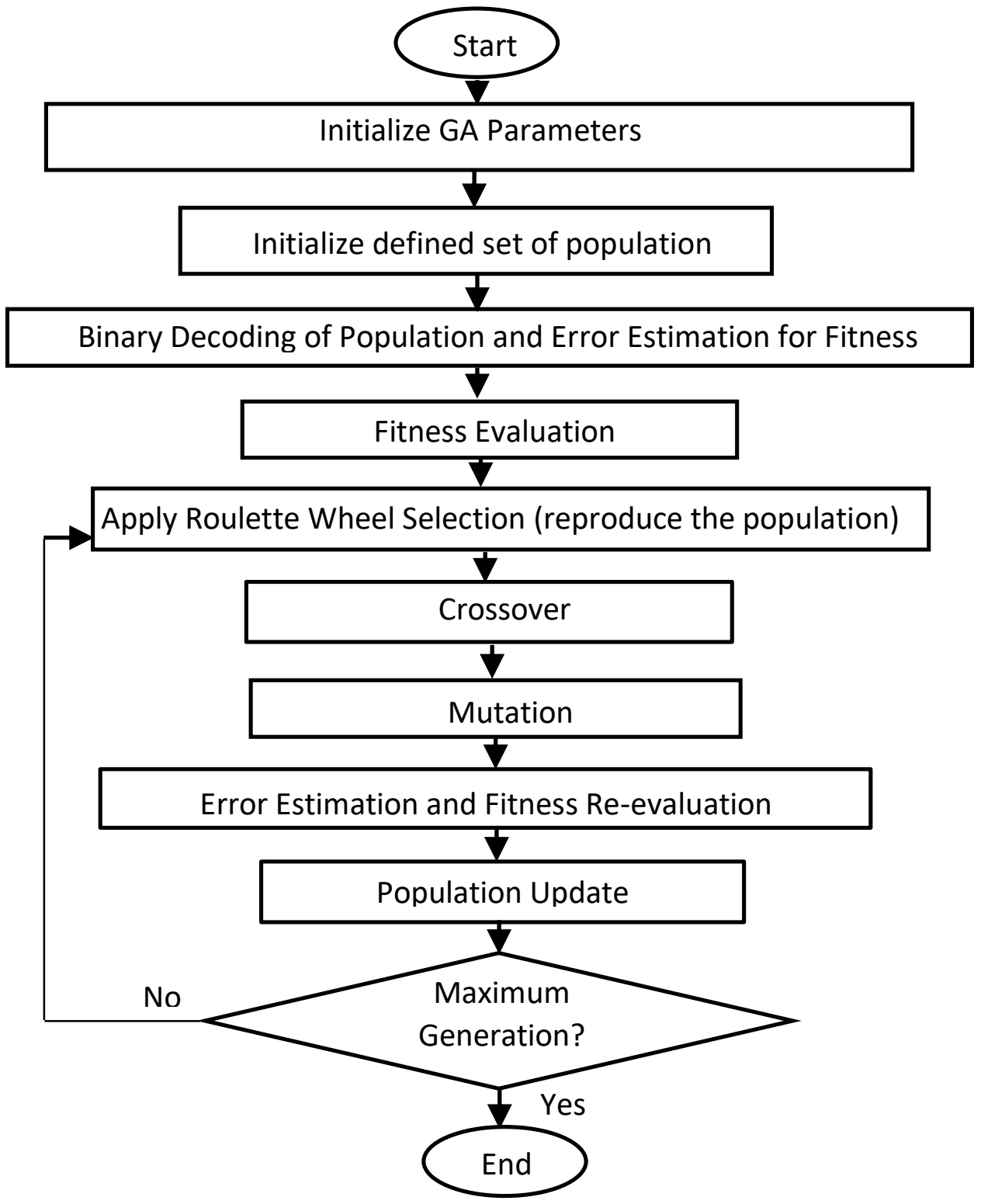

Fig. 3. Flowchart for Genetic Algorithm

The local best position at $k^{th}$ iteration is considered to be $P_{bestcost}$, and $P_{globalbest}$ is considered to be the global best position among the overall $k$ iterations. As shown in the flowchart of Fig. 4, the above processes are repeated by PSO until the maximum number of iterations are reached or a global optimum solution is found as the algorithm converges.

### *C. De-noising signals using LMS*

LMS falls under the category of gradient descent algorithms, which when initialized with an assigned value it follows the negative of gradient to locate the desired local minima of an error surface. In the case of LMS, the step size, which can be considered as the guiding factor for the algorithm, controls the negative descent to reach the local minima. The process of updating weight coefficients using the LMS can be expressed as:

$$\boldsymbol{W}[n+1] = \boldsymbol{W}[n] + \mu \boldsymbol{e}[n]\hat{\boldsymbol{Y}}[n] \qquad (9)$$

where, $\boldsymbol{W}[n]$ is the weight vector and μ is the step size. Determining the appropriate step size is found to be an important performance requirement for LMS algorithm [15]. To minimize the error signal $\boldsymbol{e}[n]$, small step size is preferred to achieve the optimal convergence speed whilst maintaining a steady performance [15]. Once the optimal weights are found, the output signal is estimated by supplying the updated weight coefficients to the filter. Fig. 5 shows the flowchart of the LMS algorithm.

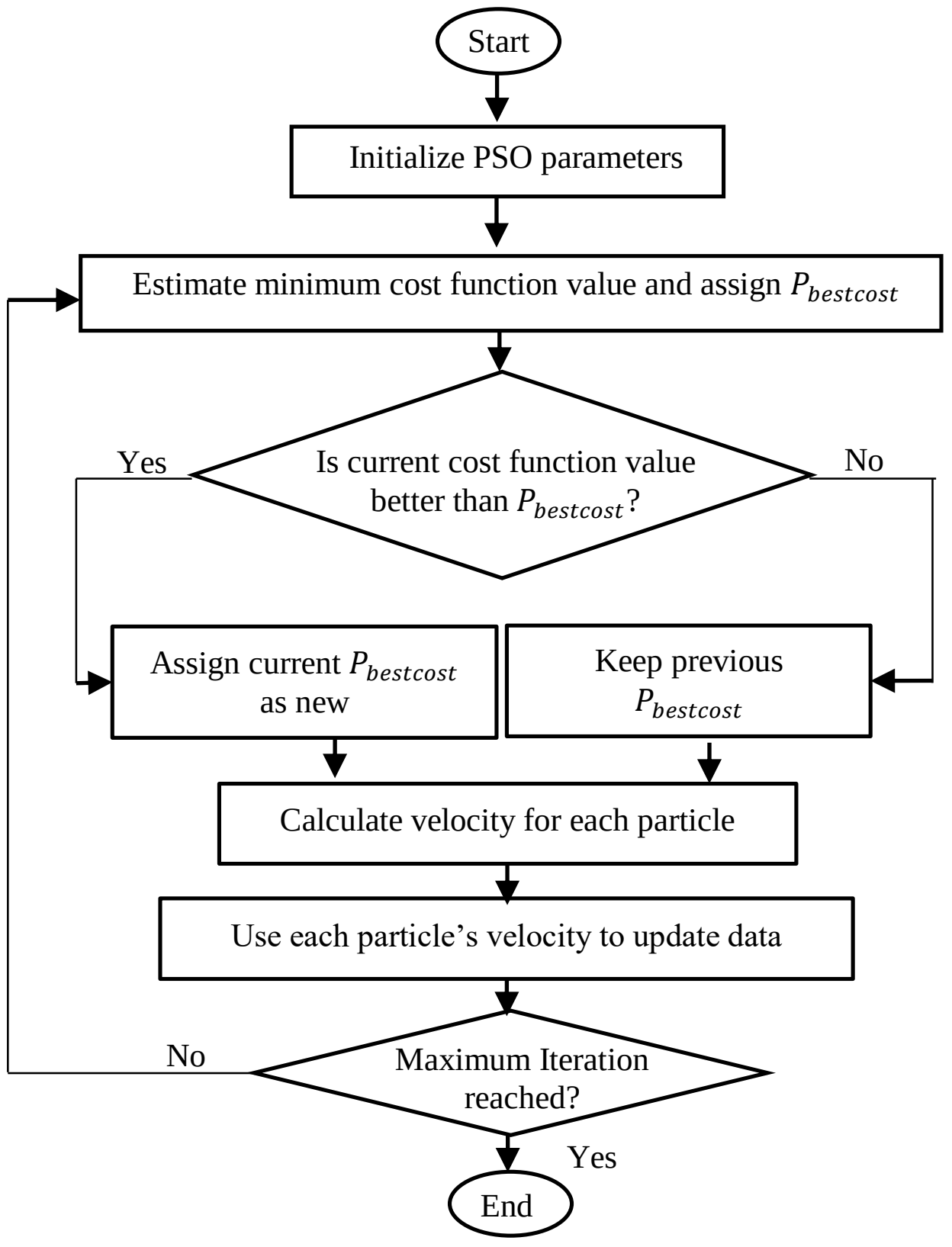

Fig. 4. Flowchart for PSO algorithm [22]

## III. Results and Comparison

The transmitted bit stream used in the simulations was generated to produce a signal of 10,000 samples, which was then modulated using M-PSK (with $M$=2) modulation scheme and transmitted over a carrier frequency of 2.4 GHz. Additive white Gaussian noise and non-linear noise were simultaneously added to the transmitted signal. At the receiver, the noisy signal was filtered using one of the three algorithms. Two metrics, bit error rate (BER) and mean square error (MSE), are used to compare the performance of these algorithms. BER, which is the ratio of bit error and total number of transmitted bits during the studied period, can be formulated as:

$$BER = \frac{Number\ of\ Corrupted\ Bits}{Total\ Number\ of\ Transmitted\ Bits} \quad (10)$$

MSE is the difference between the noisy signal and the filtered output and estimates the average of squared error. It is defined as:

$$MSE = \sum_{l=1}^{H} (Noisy\ Signal\ - Filter\ Output)^2 / H, \quad (11)$$

where, $H$ is the length of the received signal.

Fig. 6 illustrates the simulated random noisy signal generated to investigate one of the notable drawbacks of LMS. This figure shows additional noise induced frequencies besides the actual 2.4 GHz frequency. The spikes at lower frequency ranges are generated when the co-located CR antennas operate at the same time with the same frequency during full-duplex communication.

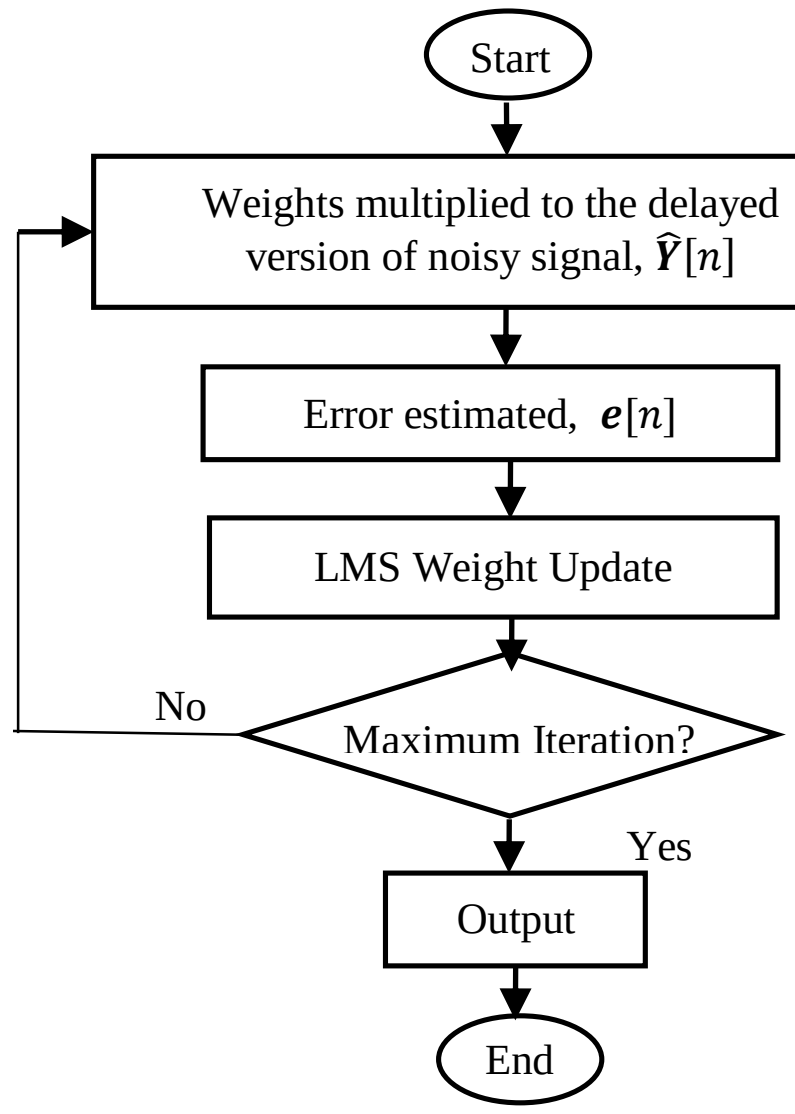

Fig. 5. Flowchart for LMS algorithm [22]

Fig. 7 shows MSE as a function of number of iterations or samples for three different step sizes of the LMS algorithm. Results were obtained using fixed -2dB SNR and filter order of L =5 for three step sizes 0.001, 0.005, and 0.01. As one can see, MSE for all step sizes increases sharply within the first 300 iterations and then gradually decreases with increasing number of iterations. MSE for step size of 0.01 is found to decrease at a faster rate and enables LMS to converge after about 5000 iterations. As step size decreases, LMS converges at a slower rate and the peak MSE increases. From these results, it can be said that the appropriate choice for step size impacts the performance of LMS.

Fig. 8 illustrates the impact of different number of population or particle sizes of GA and PSO on MSE values for both algorithms. The simulation was performed with a filter order of $L$ = 5 and 200 number of iterations/generations for a fixed SNR condition of -5dB. As observed, MSE of GA is higher than that of PSO for all the considered sizes, but is almost similar to PSO for the population sizes 60, 90, 110, 140, and 150. In addition, for the first three population sizes, MSE values for GA are higher but gradually decrease as the population size increases.

After the population size of 30, GA renders more steady MSE values with population size of 110 achieving the lowest MSE among all sizes. However, MSE for PSO remains almost constant for all the particle sizes investigated in this simulation. Therefore, for the next simulations optimal population size for GA is chosen to be 110 and for PSO particle size of 60. These optimal sizes are chosen considering factors such as computational complexity associated with iterating through large particle sizes and closest and lowest MSE values achieved by both the algorithms.

In Fig. 9, MSE over varying probability of crossover is shown for two different SNR conditions, 5dB and -5dB. The results were obtained using the population size of 110 and 300 generations for SNR -5dB and 5dB.

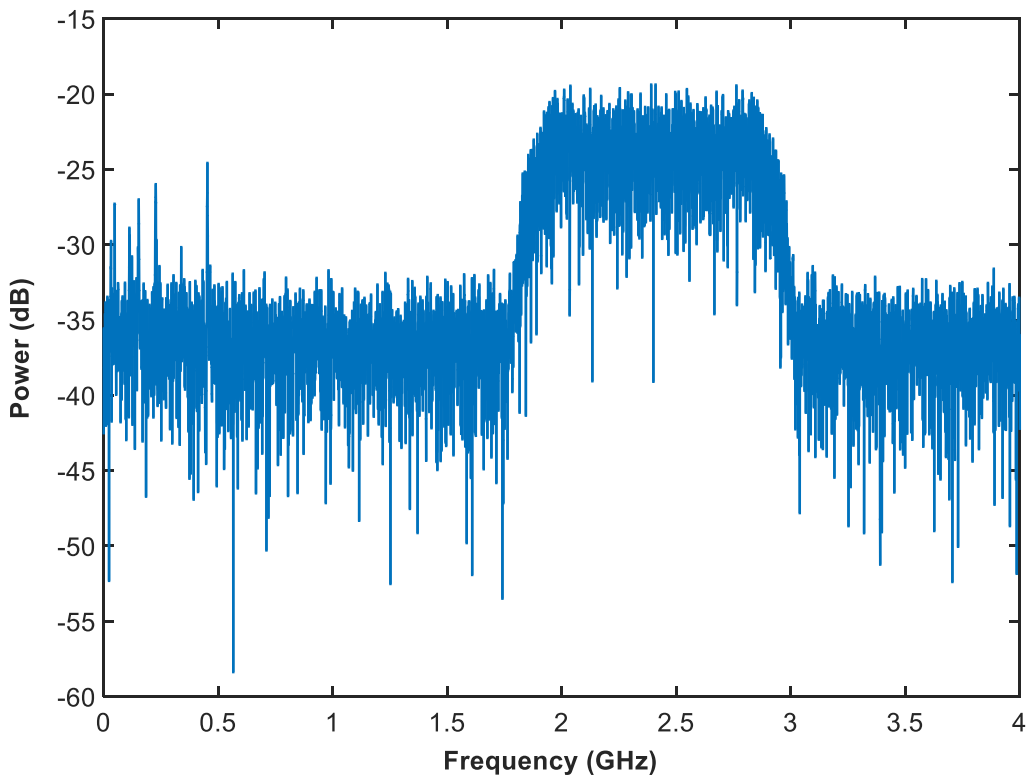

Fig 6. Received noisy signal distorted by AWGN and nonlinear noise

As one can see from this figure, MSE values do not vary for $P_c$ in the range of 0 to 0.8 under both the SNR conditions of -5dB and 5dB. As expected, MSE is higher for both $P_c$ and $P_m$ under -5dB SNR as compared to 5dB SNR. MSE for $P_c$ decreases after the probability of 0.8 for both SNR conditions and is found to be the lowest at the probability of 1. As for the probability of mutation under -5dB SNR, sharp decrease of MSE is observed in the range of 0 to 0.1 after which it does not vary significantly. For SNR conditions under 5dB. MSE of $P_m$ gradually decreases within the range of 0 to 0.3 and stabilizes for the rest of the mutation rates. Probabilities of mutation of 0.6 (under 5dB SNR) and 0.45 (under SNR -5db) are found to have the lowest MSE values.

Fig. 10 corresponds to MSE of GA, PSO, and LMS filtered signals for varying SNR conditions in the range of -10 dB to 10 dB. The simulation was performed on noisy signal distorted by AWGN and the results are obtained using a filter order L= 5, a population size of 110 for GA, optimal particle size of 60 for PSO, and a step size of 0.01 for LMS. As one can see, MSE values for all the three algorithms decrease as SNR increases, with GA and PSO having the lowest MSE values than those of LMS for all the SNR conditions. However, MSE decreases at a similar rate for GA and PSO till SNR of -2 dB. After -2dB SNR, the difference in MSE values between GA and PSO is found to increase indicating better performance of PSO than those of both GA and LMS algorithm. For all the SNR conditions, both GA and PSO outperform LMS.

Figs. 11-12 show the performances of GA, PSO, and LMS, in filtering AWGN corrupted signals and random noisy signals. BER for all algorithms are calculated to compare their performances for a range of SNR from -10 to 10 dB. Fig. 11 shows BER of AWGN distorted noisy signal under varying SNR conditions. The results are obtained using a filter order of $L$=5, step size 0.01, population of 110 for GA, and particle size of 60 for PSO. As can be seen, BER for all the algorithms decreases at a similar rate till -7 dB SNR. The difference in BER between the algorithms increases after -5dB SNR with PSO having the lowest BER followed by GA and then LMS. Both GA and PSO achieve zero BER at 1 dB and 3 dB SNR, performing significantly better than LMS.

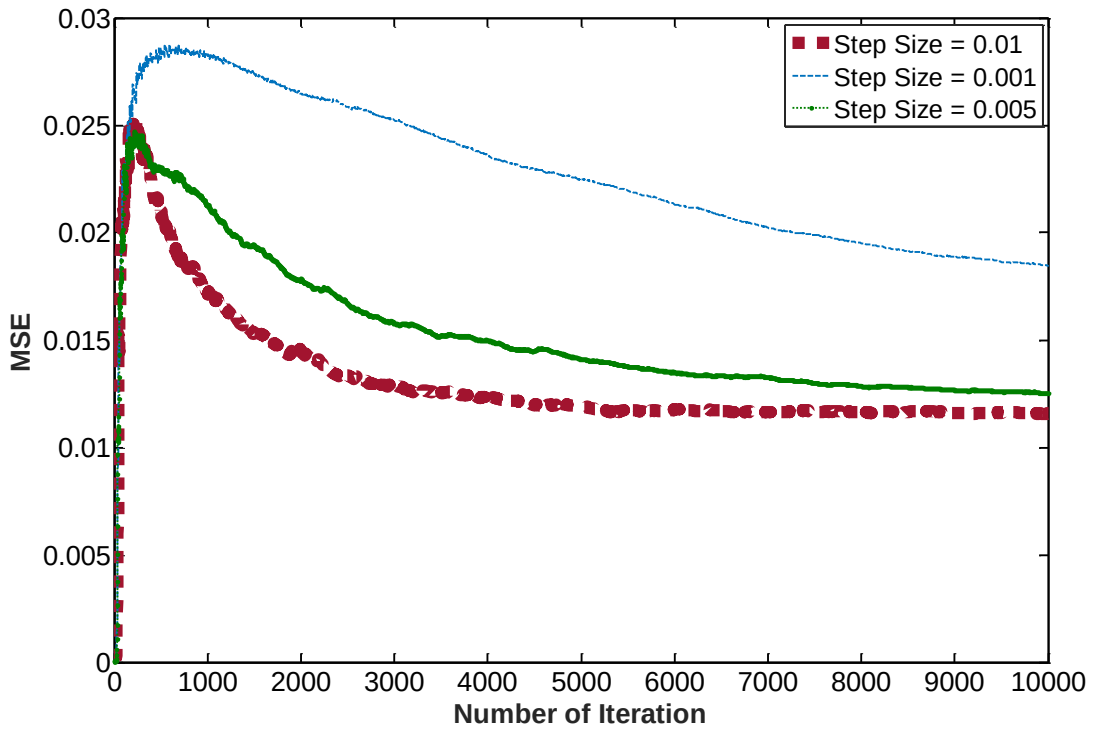

Fig. 7. Impact of *step size* on LMS convergence characteristic

From these results, it can be said that both GA and PSO are able to locate the global optimum solution for an error surface and therefore outperform LMS. However, between GA and PSO, the latter is seen to perform marginally better. But, when the number of particles or populations is considered as a performance evaluation factor, PSO performs more efficiently than GA as it requires less number of particles.

Fig. 12 shows BER of GA, PSO, and LMS filtered random noisy signals under varying SNR conditions. It can be observed that GA, PSO, and LMS perform to achieve similar BER rates under low SNR conditions in the range of -10 dB to -4 dB. After -4dB, as SNR increases GA and PSO are found to perform better than LMS.

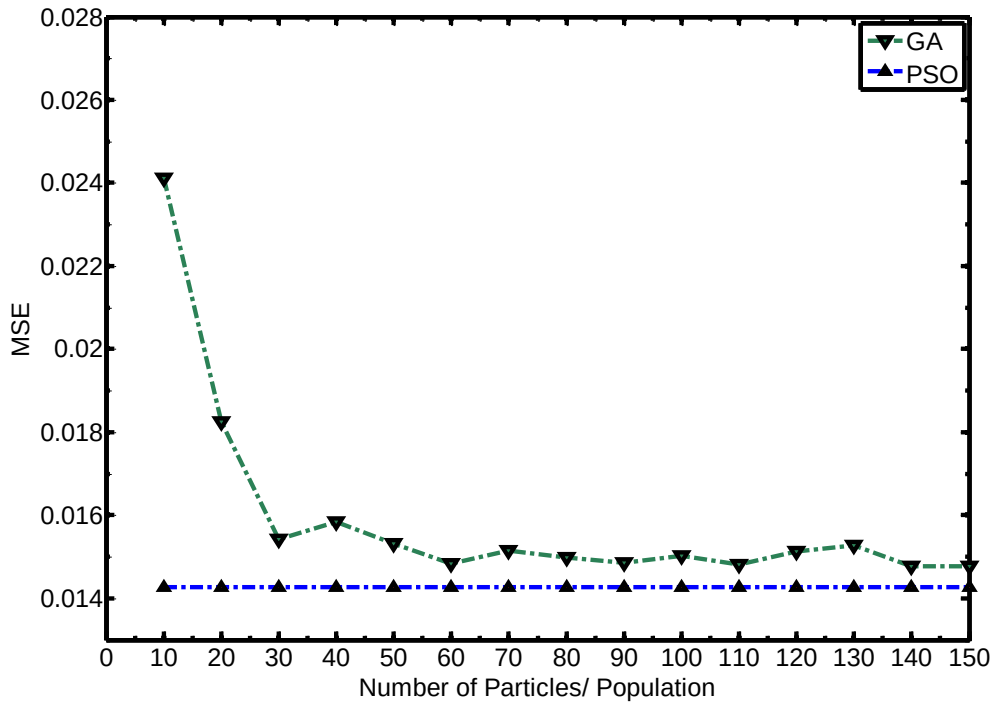

Fig. 8. MSE for different population and particle sizes of GA and PSO

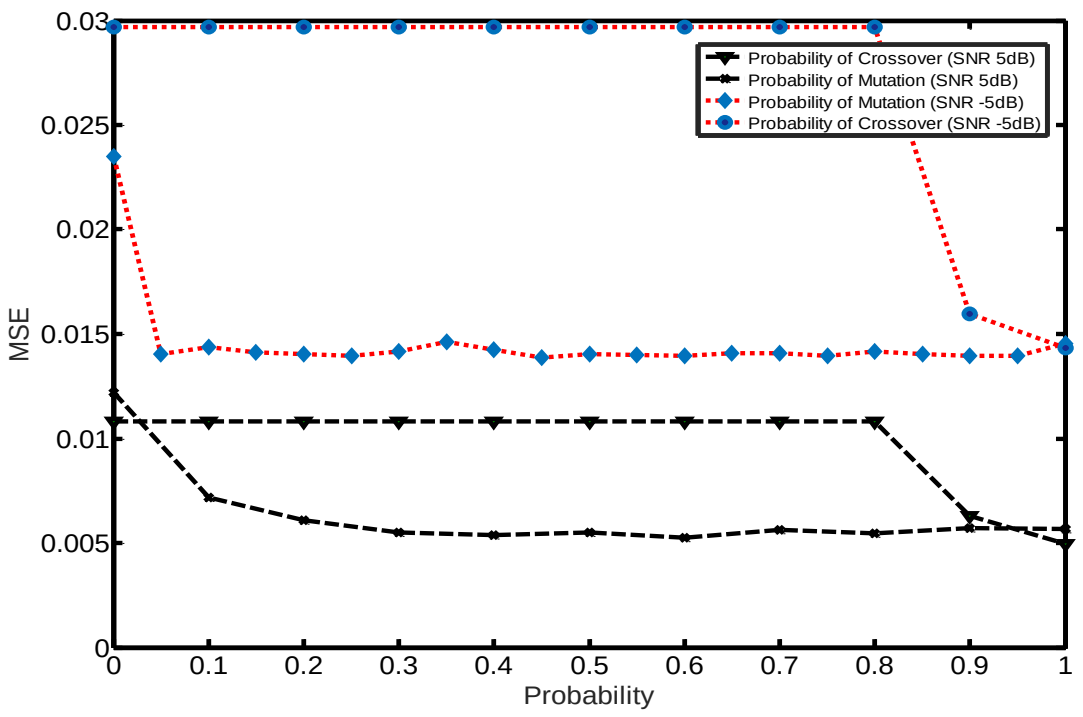

Fig. 9. Effect of Crossover and Mutation Rate of GA on MSE

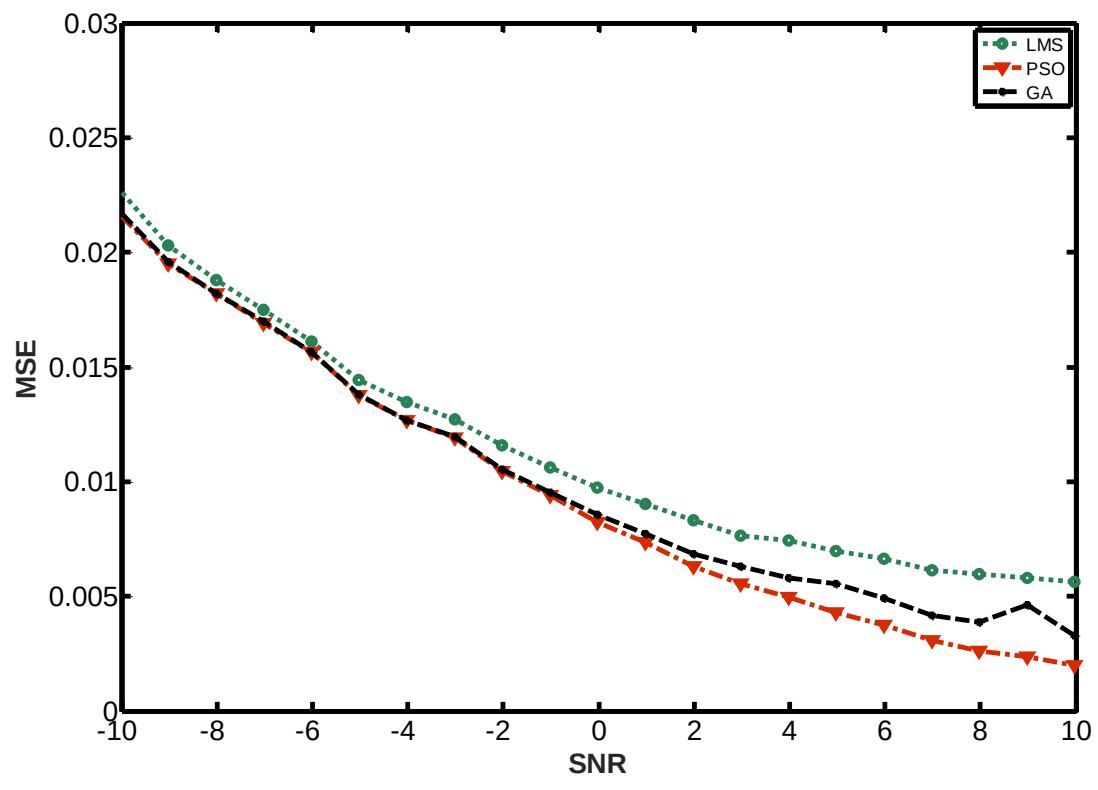

Fig. 10. MSE for GA, PSO, and LMS under varying SNR conditions

However, after 4 dB SNR, BER of GA is seen to fluctuate indicating degrading performance. Overall, PSO performs better than both GA and LMS for all the considered SNR conditions. In the case of GA, fixed control parameters (crossover and mutation rates) results in less efficient performance. The poor performance of LMS is mainly because of the increasing impact of non-linear random noise as SNR increases.

Table I outlines a general performance comparison of the 3 algorithms in terms of complexity, factors affecting their convergence rates, and optimization efficiency. In terms of computational complexity, GA and PSO are more complex than LMS. However, unlike PSO both GA and LMS require the selection of appropriate values for step size and control parameters in order to converge at an optimal rate. In terms of search optimization efficiency, GA and PSO being global optimization techniques are able to locate the global minima of a multimodal error surface. On the other hand, LMS being a local optimization technique fails to do as it can only locate local minima. Among the two global optimization techniques, GA has more steps than PSO, which increases the required processing time for GA to search for global minima. In addition, GA weight coefficients are kept in a binary-coded string format, referred to as chromosomes. These chromosomes go through crossover and mutation in every generation before they are updated. Whereas, in PSO particle position and velocity are updated at every iteration to search for the minimum cost and corresponding best solution.

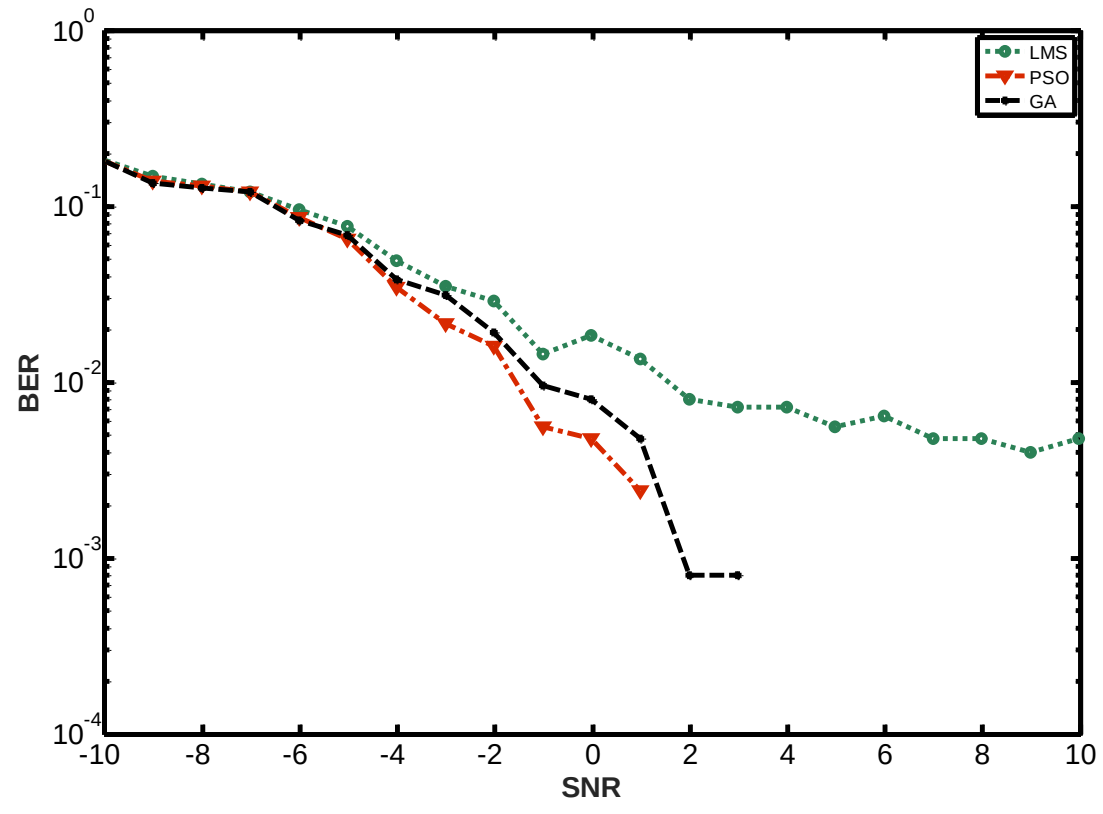

Fig. 11. BER for GA, PSO, and LMS under varying SNR conditions

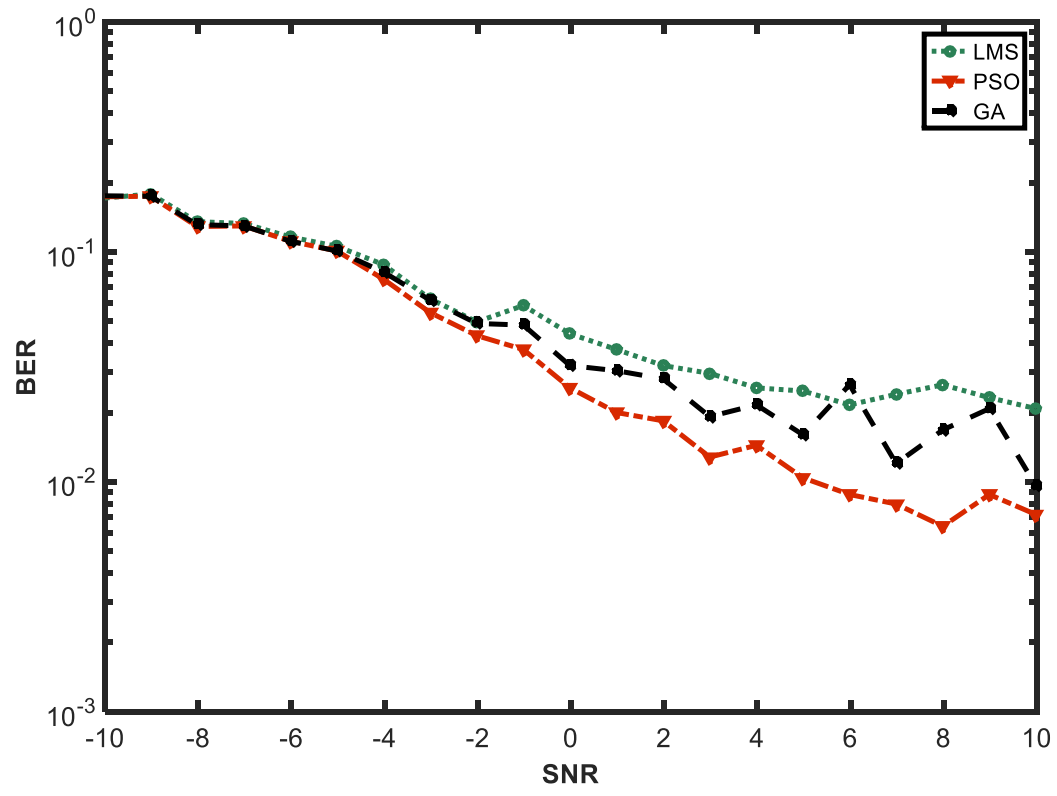

Fig. 12. BER for GA, PSO and LMS for AWGN and random noise distorted received signal

TABLE I. PERMORMANCE COMAPRISON OF THE THREE ALGORITHMS

| Algorithm | Complexity | Convergence | Optimization Efficiency |
|---|---|---|---|
| PSO | Complex | Not affected by initialization variables | • Able to identify global minima<br>• Requires less processing steps then GA |
| GA | Complex | Affected by Control Parameters e.g. – crossover and mutation rates | • Able to identify global minima<br>• Requires more processing steps and iterations than PSO |
| LMS | Simple | Affected by initialization variables, e.g. step size | Only locates local minima of error surface |

## IV. CONCLUSIONS & FUTURE WORKS

In this paper, we have described the results of a comparison study of the performances of three algorithms: GA, PSO, and LMS. Detailed simulations were performed where practical communication systems and signals were modelled with Gaussian and non-linear random noise. BER and MSE were used as performance evaluation metrics to compare the efficiencies of the three algorithms. The results show that BER values of GA and PSO are significantly better than LMS in filtering the signal distorted by Gaussian noise. However, all the three algorithms show poor performance in the case of non-linear random noise with PSO outperforming the other two algorithms. MSE for different SNR conditions were also calculated and discussed and it was shown that MSE values for GA and PSO filtered signals are lower than that of LMS. In addition to the performance metrics, population size, crossover and mutation rate for GA, and effect of particle size for PSO were also investigated.

For future work, we will implement these algorithms in Software Defined Radio (SDR) units by developing them as modules for the GNU Radio signal processing toolbox. These modules will enable real time adaptive noise cancellation. Subsequently, we will study the impact of noise due to changes in modulation scheme, transmission power, and interference through practical experiments.

## ACKNOWLEDGMENT

The work reported in this paper was supported in part by the US National Science Foundation under grant # 1443861.